\documentclass[prd,aps,showpacs,tightenlines,nofootinbib,preprint,eqsecnum]{revtex4}
\usepackage{graphicx,color,amsmath,amsxtra}
\usepackage{amssymb}
\usepackage{amsmath}

\newcommand{\bear}{\begin{array}}  \newcommand{\eear}{\end{array}}
\newcommand{\bea}{\begin{eqnarray}}  \newcommand{\eea}{\end{eqnarray}}
\newcommand{\beq}{\begin{equation}}  \newcommand{\eeq}{\end{equation}}
\newcommand{\bef}{\begin{figure}}  \newcommand{\eef}{\end{figure}}
\newcommand{\bec}{\begin{center}}  \newcommand{\eec}{\end{center}}

\newcommand{\Eqn}[1]{&\hspace{-0.2em}#1\hspace{-0.2em}&}

\def\Vec#1{\mbox{\boldmath $#1$}}

\def\Lap{{\mathop{\Delta}\limits^{(3)}}}

\def\be{\begin{equation}}
\def\ee{\end{equation}}
\def\bea{\begin{eqnarray}}
\def\eea{\end{eqnarray}}
\def\beq{\begin{eqnarray}}
\def\eeq{\end{eqnarray}}
\def\tr{{\rm tr}\, }
\def\nn{\nonumber \\}
\def\e{{\rm e}}
\def\ben{\begin{enumerate}}
\def\een{\end{enumerate}}
\def\bei{\begin{itemize}}
\def\eei{\end{itemize}}

\begin{document}

\title{
Inflationary cosmology and the late-time accelerated expansion \\
of the universe in non-minimal Yang-Mills-$F(R)$ gravity \\
and non-minimal vector-$F(R)$ gravity
}

\author{Kazuharu Bamba$^1$\footnote{
Present address: 
Department of Physics, National Tsing Hua University, 
Hsinchu, Taiwan 300}, Shin'ichi Nojiri$^2$
and Sergei D. Odintsov$^3$\footnote{
also at Lab. Fundam. Study, Tomsk State
Pedagogical University, Tomsk}}
\affiliation{
$^1$Department of Physics, Kinki University, Higashi-Osaka 577-8502, Japan\\
$^2$Department of Physics, Nagoya University, Nagoya 464-8602. Japan\\
$^3$Instituci\`{o} Catalana de Recerca i Estudis Avan\c{c}ats (ICREA)
and Institut de Ciencies de l'Espai (IEEC-CSIC),
Campus UAB, Facultat de Ciencies, Torre C5-Par-2a pl, E-08193 Bellaterra
(Barcelona), Spain
}


\begin{abstract}
We study inflationary cosmology and the late-time accelerated expansion of
the universe in non-minimal Yang-Mills (YM) theory,
in which the YM field couples to a function of the scalar curvature.
It is shown that power-law inflation can be realized due to the non-minimal
YM field-gravitational coupling  which maybe caused by quantum
corrections.
Moreover, it is demonstrated that both inflation and the late-time
accelerated expansion of the universe can be realized in a modified
YM-$F(R)$ gravity which is consistent with solar system tests. 
Furthermore, it is shown that this result can be realized also
in a non-minimal vector-$F(R)$ gravity.
In addition, we consider the duality of the non-minimal electromagnetic
theory and that of the non-minimal YM theory,
and also discuss the cosmological reconstruction of the YM theory.
\end{abstract}


\maketitle

\section{Introduction}

Recent observations confirmed that there existed the inflationary stage
in the early universe, and that also at present the expansion of the universe
is accelerating~\cite{WMAP1, SN1}.
Various scenarios for the late-time acceleration in the expansion of the
universe has been proposed. In fact, however, the cosmic acceleration
mechanism
is not well understood yet
(for recent reviews, see~\cite{Peebles:2002gy, Padmanabhan:2002ji,
Copeland:2006wr, Durrer:2007re, NO-rev}).

There exists two approaches to account for the late-time acceleration
of the universe.
One is dark energy, i.e., general relativistic approach.
The other is dark gravity, i.e., modified gravity approach.
Among the latter approaches studied so far,
the modifications to the Einstein-Hilbert action, e.g.,
the addition of an arbitrary function of the scalar curvature to it,
is one of the most promising latter approaches
(for a review, see~\cite{NO-rev}).
Such a modified theory must pass cosmological bounds and solar system tests
because it is considered as an alternative gravitational theory.

A very realistic modified gravitational theory that evade
solar-system tests has recently been proposed
by Hu and Sawicki~\cite{Hu:2007nk}
(for related studies, see~\cite{related studies}).
In this theory, an effective epoch described by the cold dark matter model
with cosmological constant ($\Lambda$CDM), which accounts for high-precision
observational data, is realized as in general relativity with
cosmological constant (for a review of observational data confronted with
modified gravity, see~\cite{Capozziello:2007ec}).
This theory can successfully explain the late-time acceleration
of the universe. In Ref.~\cite{Hu:2007nk}, however,
the possibility of the realization of inflation has not been discussed.
In Refs.~\cite{Nojiri:2007as,NO1,Eli}, therefore,
modified gravities in which both inflation and the late-time acceleration
of the universe can be realized,
following the previous inflation-acceleration
unification proposal~\cite{Nojiri:2003ft}, have been presented and
investigated. The classification of viable $F(R)$ gravities has also been
suggested in Ref.~\cite{Eli}.
Here, $F(R)$ is an arbitrary function of the scalar curvature $R$.

Furthermore, there exists another gravitational source of inflation and the
late-time acceleration of the universe: a coupling between the scalar
curvature and matter Lagrangian~\cite{matter-1, Allemandi:2005qs} 
(see also~\cite{Deruelle:2008fs}).
Such a coupling may be applied for the realization of the dynamical
cancellation of cosmological constant~\cite{DC}.
In Refs.~\cite{criteria-1, Faraoni:2007sn, Bertolami:2007vu},
the criteria for the viability of such theories have been considered.
As a simple case, a coupling between a function of
the scalar curvature and the kinetic term of a massless scalar field
in a viable modified gravity has been considered~\cite{Nojiri:2007bt}.

Recently,
inflation and the late-time acceleration of the universe in non-minimal
electromagnetism, in which the electromagnetic field couples to a function of
the scalar curvature, have been studied in Ref.~\cite{Bamba:2008ja} by 
using the analyzing procedure in the electromagnetic field considered in 
Ref.~\cite{Bamba-mag-2}. 
It is known that the coupling between the scalar
curvature and the Lagrangian of the electromagnetic field arises in curved
spacetime due to one-loop vacuum-polarization effects in
Quantum Electrodynamics (QED)~\cite{Drummond:1979pp}.
As a result, it has been shown that power-law inflation
can be realized due to the non-minimal gravitational coupling of the
electromagnetic field, and that large-scale magnetic fields can be generated
due to the breaking of the conformal invariance of the electromagnetic field
through its non-minimal gravitational coupling\footnote{
In Ref.~\cite{G-M inflation from 5D},
gravitational-electromagnetic inflation from a 5-dimensional vacuum state
has been considered.} (see also~\cite{Campanelli:2008qp}).
The mechanism of inflation in this model is as follows.
In the very early universe before inflation, electromagnetic quantum
fluctuations are generated due to the breaking of the conformal invariance of
the electromagnetic field and they act as a source for inflation.
Furthermore, also during inflation electromagnetic quantum
fluctuations are newly generated and the scale is stretched due to inflation,
so that the scale can be larger than the Hubble horizon at that time,
and they lead to the large-scale magnetic fields observed in galaxies and
clusters of galaxies.
This idea is based on the assumption that
a given mode is excited quantum mechanically while it is subhorizon sized and
then as it crosses outside the horizon ``freezes in'' as a classical
fluctuation~\cite{Turner:1987bw}.
These large-scale magnetic fields can be the origin of the large-scale
magnetic fields
with the field strength $10^{-7}$--$10^{-6}$G on 10kpc--1Mpc
scale observed in clusters of galaxies~\cite{clusters of galaxies}
(for reviews of cosmic magnetic fields, see~ \cite{magnetic-field-review}).
Furthermore, it has been demonstrated that both inflation and
the late-time acceleration of the universe can be realized in a modified
Maxwell-$F(R)$ gravity proposed in Ref.~\cite{Nojiri:2007as} 
which is consistent with solar system tests. 

In the present paper, we consider inflationary cosmology and the late-time
accelerated expansion of the universe in non-minimal non-Abelian gauge theory,
called the Yang-Mills (YM) theory, in which the non-Abelian gauge field
(the YM field) couples to a function of the scalar curvature, 
in order to investigate the cosmological consequences of the non-minimal 
gravitational coupling of the YM filed. 
Furthermore, we consider a non-minimal vector-$F(R)$ gravity.
In the past studies, inflation driven by a vector filed has been
discussed~\cite{Ford:1989me, Golovnev:2008cf}.
Moreover, as a candidate for dark energy,
the effective YM condensate~\cite{YM-DE-1, YM-DE-2},
the Born-Infeld quantum condensate~\cite{Elizalde:2003ku}
and a vector field~\cite{ArmendarizPicon:2004pm, Wei:2006tn, Wei:2006gv,
Jimenez:2008au, KM1} have been proposed.
In particular, the possibility that
the accelerated expansion of the universe is driven by a field with
an anisotropic equation of state has been considered in Ref.~\cite{KM1}.
As a result, 
we show that power-law inflation can be realized due to the non-minimal
gravitational coupling of the YM field\footnote{
In Ref.~\cite{chromomagnetic fields},
the spontaneous generation of chromomagnetic fields at
high temperature has been investigated.}. 
Moreover, we demonstrate that both inflation and the late-time 
accelerated expansion of the universe can be realized in a modified
Yang-Mills-$F(R)$ gravity which is consistent with solar system tests. 
Furthermore, we show that this result can be realized also in
a non-minimal vector-$F(R)$ gravity. 
In addition, we consider the duality of the non-minimal electromagnetic
theory and that of the non-minimal YM theory,
and also discuss the reconstruction of the YM theory.

There are several motivations to study non-minimal YM theory. 
First of all, we show that the appearance of such non-minimal terms in the 
early universe is compatible with current formulations of YM theory due to 
specific choice of non-minimal function. 
Second, some string compactification may lead to effective 
scalars-YM-Einstein theory (plus higher order corrections). In some cases, 
one can delete scalars in such a way, that extra curvature terms 
(non-minimal ones) appear in front of YM Lagrangian. 
Third, the celebrated asymptotic freedom phenomenon maybe understood as 
appearance of non-minimal terms at the early universe. 

This paper is organized as follows.
In Sec.\ II we consider a non-minimal gravitational coupling of
the $SU(N)$ YM field in general relativity.
First, we describe our model and derive equations of motion from it.
Next, 
we analyze the gravitational field equation, and then show that power-law 
inflation can be realized.
In Sec.\ III we consider a non-minimal gravitational coupling of the 
$SU(N)$ YM field in a modified gravitational theory proposed
in Ref.~\cite{Nojiri:2007as}. We show that in this theory both inflation and
the late-time acceleration of the universe can be realized.
In Sec.\ IV we consider a non-minimal vector-$F(R)$ gravity.
Furthermore,
in Sec.\ V we consider the duality of the non-minimal electromagnetic
theory and that of the non-minimal YM theory.
In addition, in Sec.\ VI we discuss the reconstruction of the YM theory.
Finally, summary is given in Sec.\ VII.
We use units in which $k_\mathrm{B} = c = \hbar = 1$ and denote the
gravitational constant $8 \pi G$ by ${\kappa}^2$, so that
${\kappa}^2 \equiv 8\pi/{M_{\mathrm{Pl}}}^2$, where
$M_{\mathrm{Pl}} = G^{-1/2} = 1.2 \times 10^{19}$GeV is the Planck mass.
Moreover, in terms of electromagnetism we adopt Heaviside-Lorentz units.

\section{Inflation in general relativity}

In this section, following the discussion given in Ref.~\cite{Bamba:2008ja}.
we first consider a non-minimal gravitational coupling of the YM field in
general relativity.

\subsection{Model}

We consider the following model action:
\begin{eqnarray}
S_{\mathrm{GR}} \Eqn{=}
\int d^{4}x \sqrt{-g}
\left[ \hspace{1mm}
{\mathcal{L}}_{\mathrm{EH}}
+{\mathcal{L}}_{\mathrm{YM}}
\hspace{1mm} \right]\,,
\label{eq:2.1} \\[2mm]
{\mathcal{L}}_{\mathrm{EH}}
\Eqn{=}
\frac{1}{2\kappa^2} R\,,
\label{eq:2.2} \\[2mm]
{\mathcal{L}}_{\mathrm{YM}}
\Eqn{=}
-\frac{1}{4} I(R) F_{\mu\nu}^{a}F^{a\mu\nu}
\left[1+b\tilde{g}^2 \ln
\left| \frac{-\left( 1/2 \right) F_{\mu\nu}^{a}F^{a\mu\nu}}{\mu^4} \right|
\right]\,,
\label{eq:2.3}
\end{eqnarray}
with
\begin{eqnarray}
I(R) \Eqn{=} 1+f(R)\,,
\label{eq:2.4} \\[2mm]
b \Eqn{=} \frac{1}{4} \frac{1}{8 \pi^2}\frac{11}{3}N\,,
\label{eq:2.5} \\[2mm]
F_{\mu\nu}^{a} \Eqn{=}
{\partial}_{\mu}A_{\nu}^{a} - {\partial}_{\nu}A_{\mu}^{a}
+ f^{abc} A_{\mu}^{b} A_{\nu}^{c}\,,
\label{eq:2.6}
\end{eqnarray}
where $g$ is the determinant of the metric tensor $g_{\mu\nu}$,
$R$ is the scalar curvature arising from the spacetime
metric tensor $g_{\mu\nu}$,
and ${\mathcal{L}}_{\mathrm{EH}}$ is the Einstein-Hilbert action.
Moreover, ${\mathcal{L}}_{\mathrm{YM}}$ with $I(R)=1$ is the effective
Lagrangian of the $SU(N)$ YM theory up to one-loop
order~\cite{YM-L, Adler:1983zh},
$f(R)$ is an arbitrary function of $R$,
$b$ is the asymptotic freedom constant,
$F_{\mu\nu}^{a}$ is the field strength tensor,
$A_{\mu}^{a}$ is the $SU(N)$ YM field
with the internal symmetry index $a$
(Roman indices, $a$, $b$, $c$, run over $1, 2, \ldots, N^2-1$, and
in $F_{\mu\nu}^{a}F^{a\mu\nu}$ the summation in terms of the index $a$ is
also made), and
$f^{abc}$ is a set of numbers called structure constants and completely
antisymmetric~\cite{P-S}.
Furthermore, $\mu$ is the mass scale of the renormalization point, and
a field-strength-dependent running coupling constant
is given by~\cite{Adler:1983zh}
\begin{eqnarray}
\tilde{g}^2 (X)
= \frac{\tilde{g}^2}{1+b\tilde{g}^2\ln \left| X/ \mu^4 \right|}\,,
\label{eq:2.7}
\end{eqnarray}
where
\begin{eqnarray}
X \equiv -\frac{1}{2} F_{\mu\nu}^{a} F^{a\mu\nu}\,.
\label{eq:2.8}
\end{eqnarray}
Hence, $\tilde{g}$ is the value of the running coupling constant
when $X = \mu^4$.

The field equations can be derived by taking variations of the
action in Eq.~(\ref{eq:2.1}) with respect to the
metric $g_{\mu\nu}$ and the $SU(N)$ YM field $A_{\mu}^{a}$ as follows:
\begin{eqnarray}
R_{\mu \nu} - \frac{1}{2}g_{\mu \nu}R
= \kappa^2 T^{(\mathrm{YM})}_{\mu \nu}\,,
\label{eq:2.9}
\end{eqnarray}
with
\begin{eqnarray}
\hspace{-5mm}
T^{(\mathrm{YM})}_{\mu \nu}
\Eqn{=}
I(R) \left( g^{\alpha\beta} F_{\mu\beta}^{a} F_{\nu\alpha}^{a} \varepsilon
-\frac{1}{4} g_{\mu\nu} \mathcal{F} \right)
\nonumber \\[2mm]
&&
{}+\frac{1}{2} \left\{ f^{\prime}(R) \mathcal{F} R_{\mu \nu}
+ g_{\mu \nu} \Box \left[ f^{\prime}(R) \mathcal{F} \right]
- {\nabla}_{\mu} {\nabla}_{\nu}
\left[ f^{\prime}(R) \mathcal{F} \right]
\right\}\,,
\label{eq:2.10} \\[2mm]
\varepsilon \Eqn{=}
1+b\tilde{g}^2 \ln \left| e
\left[ \frac{-\left( 1/2 \right) F_{\mu\nu}^{a}F^{a\mu\nu}}{\mu^4}
\right] \right|
= 1+b\tilde{g}^2 \ln \left| e \left( \frac{X}{\mu^4} \right)
\right|\,,
\label{eq:2.11} \\[2mm]
\mathcal{F} \Eqn{=}
F_{\mu\nu}^{a}F^{a\mu\nu}
\left[ 1+b\tilde{g}^2 \ln \left|
\frac{-\left( 1/2 \right) F_{\mu\nu}^{a}F^{a\mu\nu}}{\mu^4} \right|
\right]
= -2X\left( 1+b\tilde{g}^2 \ln \left| \frac{X}{\mu^4} \right| \right)\,,
\label{eq:2.12}
\end{eqnarray}
and
\begin{eqnarray}
\frac{1}{\sqrt{-g}}{\partial}_{\mu}
\left[ \sqrt{-g} I(R) \varepsilon
F^{a\mu\nu} \right]
-I(R) \varepsilon f^{abc} A_{\mu}^{b} F^{c\mu\nu}
= 0\,,
\label{eq:2.13}
\end{eqnarray}
where the prime denotes differentiation with respect to $R$,
${\nabla}_{\mu}$ is the covariant derivative operator associated with
$g_{\mu \nu}$, and $\Box \equiv g^{\mu \nu} {\nabla}_{\mu} {\nabla}_{\nu}$
is the covariant d'Alembertian for a scalar field.
In addition, $R_{\mu \nu}$ is the Ricci curvature tensor,
while $T^{(\mathrm{YM})}_{\mu \nu}$ is the contribution to
the energy-momentum tensor from the $SU(N)$ YM field.
Moreover, $\varepsilon$ is a field-strength-dependent effective dielectric
constant~\cite{Adler:1983zh}, and $e \approx 2.72$ is the Napierian number.
In deriving the second equalities in
Eqs.\ (\ref{eq:2.11}) and (\ref{eq:2.12}), we have used Eq.\ (\ref{eq:2.8}).

We assume the spatially flat
Friedmann-Robertson-Walker (FRW) spacetime with the metric
\begin{eqnarray}
{ds}^2 =-{dt}^2 + a^2(t)d{\Vec{x}}^2
= a^2(\eta) ( -{d \eta}^2 + d{\Vec{x}}^2 )\,,
\label{eq:2.14}
\end{eqnarray}
where $a$ is the scale factor, and $\eta$ is the conformal time.
In this spacetime,
$g_{\mu \nu} = \mathrm{diag} \left(-1, a^2(t), a^2(t), a^2(t) \right)$, and
the components of $R_{\mu \nu}$ and $R$ are given by
\begin{eqnarray}
R_{00} = -3\left( \dot{H} + H^2 \right)\,,
\hspace{1mm}
R_{0i} = 0\,,
\hspace{1mm}
R_{ij} = \left( \dot{H} + 3H^2 \right) g_{ij}\,,
\hspace{1mm}
R=6\left( \dot{H} + 2H^2 \right)\,,
\label{eq:2.15}
\end{eqnarray}
where $H=\dot a/a$ is the Hubble parameter. Here, a dot denotes a time
derivative, $\dot{~}=\partial/\partial t$.

\subsection{Power-law inflation}

The $(\mu,\nu)=(0,0)$ component and
the trace part of the $(\mu,\nu)=(i,j)$ component of Eq.~(\ref{eq:2.9}),
where $i$ and $j$ run from $1$ to $3$, read
\begin{eqnarray}
H^2 +J_1 
\Eqn{=} 
\frac{\kappa^2}{6}
\biggl\{
I(R) \left( b\tilde{g}^2 X + \varepsilon Y \right)
+ 3 \left[
-f^{\prime}(R) \left( \dot{H} + H^2 \right) +
6 f^{\prime\prime}(R) H \left( \ddot{H} + 4H\dot{H} \right)
\right] \mathcal{F}
\nonumber \\[2mm]
&&
{}+ 3 f^{\prime}(R) H \dot{\mathcal{F}}
\biggr\}\,,
\label{eq:2.38} 
\end{eqnarray}
\begin{eqnarray}
J_1 
= 
\frac{1}{6} F(R)  - F^{\prime}(R) \left( \dot{H} + H^2 \right)\,,
\label{eq:2.39}
\end{eqnarray}
and
\begin{eqnarray}
\hspace{-12mm}
2\dot{H} + 3H^2 +J_2
\Eqn{=}
\frac{\kappa^2}{2}
\biggl\{
I(R) X \left( - \frac{1}{3} \varepsilon + b\tilde{g}^2 \right)
+ \biggl[
-f^{\prime}(R) \left( \dot{H} + 3H^2 \right)
\nonumber \\[2mm]
&& \hspace{-8mm}
{}+6f^{\prime\prime}(R)
\left(\dddot{H}+7H\ddot{H}+4\dot{H}^2+12H^2\dot{H} \right)
+36f^{\prime\prime\prime}(R) \left( \ddot{H}+4H\dot{H} \right)^2
\biggr] \mathcal{F}
\nonumber \\[2mm]
&& \hspace{-8mm}
{}+3\left[f^{\prime}(R)H + 4f^{\prime\prime}(R)
\left( \ddot{H}+4H\dot{H} \right) \right] \dot{\mathcal{F}}
+ f^{\prime}(R) \ddot{\mathcal{F}}
\biggr\}\,,
\label{eq:2.40} \\[2mm]
J_2
\Eqn{=}
\frac{1}{2}F(R)
- F^{\prime}(R) \left( \dot{H} + 3H^2 \right)
+ 6F^{\prime\prime}(R) \left[
\dddot{H} + 4\left( \dot{H}^2 + H\ddot{H} \right)
\right]
\nonumber \\[2mm]
&& \hspace{0mm}
{}+ 36F^{\prime\prime\prime}(R) \left( \ddot{H} + 4H\dot{H} \right)^2\,,
\label{eq:2.41}
\end{eqnarray}
respectively.
Here, $X$ and $Y$ are given by
\begin{eqnarray}
X \Eqn{=}
|{E_i^{a}}^{(\mathrm{proper})}(t)|^2
- |{B_i^{a}}^{(\mathrm{proper})}(t)|^2\,,
\label{eq:2.42} \\[2mm]
Y \Eqn{=}
|{E_i^{a}}^{(\mathrm{proper})}(t)|^2
+ |{B_i^{a}}^{(\mathrm{proper})}(t)|^2\,, 
\label{eq:2.43}
\end{eqnarray}
where ${E_i^{a}}^{(\mathrm{proper})}(t)$ and 
${B_i^{a}}^{(\mathrm{proper})}(t)$ are 
the quantities corresponding to proper electric and magnetic fields
in the $SU(N)$ YM theory, respectively. 
In this paper, because we consider the case in which there exist 
the YM electric and magnetic fields as background quantities 
at the 0th order, we here consider that the YM electric and magnetic fields 
do not have the dependence on the space components $\Vec{x}$. 
Moreover, $J_1$ and $J_2$ are correction terms in
a modified gravitational theory described by the action
in Eq.~(\ref{eq:3.1}) in
the next section. Hence, because in this section we consider
general relativity, i.e., the case $F(R)=0$ in the action
in Eq.~(\ref{eq:3.2}),
here both $J_1$ and $J_2$ are zero.
Furthermore, 
in deriving Eqs.~(\ref{eq:2.38}) and (\ref{eq:2.40}), we have used 
equations in (\ref{eq:2.15}) and the following equation:
\begin{eqnarray}
\mathcal{F}=-2X(\varepsilon - b\tilde{g}^2)\,,
\label{eq:2.44}
\end{eqnarray}
which follows from
Eqs.~(\ref{eq:2.8}), (\ref{eq:2.11}) and (\ref{eq:2.12}).

In the search of exact solutions for non-minimal YM (electromagnetic)-gravity 
theory (see~\cite{Balakin:2005fu, Balakin:2007nw}), 
the problem of off-diagonal components of YM (electromagnetic) stress tensor 
being non-zero while the right-hand side of Einstein equations is 
zero. In our case, we consider as follows. 
As a simple case, we can consider the following case in which 
the off-diagonal components of $T^{(\mathrm{YM})}_{\mu \nu}$ 
in Eq.~(\ref{eq:2.10}) vanishes: 
(i) Only (YM) magnetic fields are generated and hence (YM) electric fields 
are negligible. 
(ii) $\Vec{B^{a}} = (B^{a}_1, B^{a}_2, B^{a}_3)$, where 
$B^{a}_1 = B^{a}_2 =0, B^{a}_3 \neq 0$, namely,
we consider the case in which only one component of $\Vec{B}^{a}$ is non-zero and hence other two components are zero.
In such a case, it follows from $\mathrm{div} \Vec{B}^{a} = 0$ that 
the off-diagonal components of the last term 
on the right-hand side of 
$T^{(\mathrm{YM})}_{\mu \nu}$, i.e., 
${\nabla}_{\mu} {\nabla}_{\nu} \left[ f^{\prime}(R) \mathcal{F} \right]$
are zero. 
Thus, all of the off-diagonal components of $T^{(\mathrm{YM})}_{\mu \nu}$ 
are zero. 
In this paper (including Secs.~III and IV) we consider the above case in order 
to investigate the cosmological consequences of the non-minimal gravitational 
coupling of the YM filed. 

In Eq.\ (\ref{eq:2.13}), because the amplitude of $A_{\mu}^{a}$ is small, 
we can neglect the higher order than or equal to the quadratic terms in 
$A_{\mu}^{a}$ and investigate the linearized equation of 
Eq.\ (\ref{eq:2.13}) in terms of $A_{\mu}^{a}$. 
The linearized equation of motion in the Coulomb gauge,
${\partial}^jA_j^{a}(t,\Vec{x}) =0$, 
and the case of $A_{0}^{a}(t,\Vec{x}) = 0$, reads
\begin{eqnarray}
{\ddot{A}}_i^{a}(t,\Vec{x})
+ \left( H + \frac{\dot{I}}{I}
\right) {\dot{A}}_i^{a}(t,\Vec{x})
- \frac{1}{a^2}\Lap\, A_i^{a}(t,\Vec{x}) = 0\,,
\label{eq:A-1}
\end{eqnarray}
where $\Lap =  {\partial}^i {\partial}_i$ is the flat 3-dimensional Laplacian. 
It follows from Eq.~(\ref{eq:A-1}) that the Fourier mode $A_i^{a}(k,t)$ 
satisfies the equation 
\begin{eqnarray}
{\ddot{A}}_i^{a}(k,t) + \left( H + \frac{\dot{I}}{I} \right)
               {\dot{A}}_i^{a}(k,t) + \frac{k^2}{a^2} A_i^{a}(k,t) = 0\,. 
\label{eq:A-2}
\end{eqnarray}
Replacing the independent variable $t$ by $\eta$, we find that
Eq.~(\ref{eq:A-2}) becomes
\begin{eqnarray}
\frac{\partial^2 A_i^{a}(k,\eta)}{\partial \eta^2} +
\frac{1}{I(\eta)} \frac{d I(\eta)}{d \eta}
\frac{\partial A_i^{a}(k,\eta)}{\partial \eta}
+ k^2 A_i^{a}(k,\eta) = 0\,.
\label{eq:A-3}
\end{eqnarray}
By using the WKB approximation on subhorizon scales and the long-wavelength 
approximation on superhorizon scales, and matching these solutions at the 
horizon crossing~\cite{Bamba-mag-2}, we find 
\begin{eqnarray}
\hspace{-5mm}
\left|A_i^{a}(k,\eta)\right|^2
= |C(k)|^2
= \frac{1}{2kI(\eta_k)}
\left|1- \left[ \frac{1}{2}\frac{1}{kI(\eta_k)}\frac{d I(\eta_k)}{d \eta}
+ i \right]k\int_{\eta_k}^{{\eta}_{\mathrm{f}}}
\frac{I(\eta_k)}{I \left(\Tilde{\Tilde{\eta}} \right)}
d\Tilde{\Tilde{\eta}}\,\right|^2\,,
\label{eq:A-4}
\end{eqnarray}
where $\eta_k$ and ${\eta}_{\mathrm{f}}$ are the conformal time 
at the horizon-crossing and one at the end of inflation, respectively. 
Consequently, from Eq.~(\ref{eq:A-4}) we obtain the amplitude of 
the proper YM magnetic fields on a comoving scale $L=2\pi/k$ 
in the position space 
\begin{eqnarray}
|{B_i^{a}}^{(\mathrm{proper})}(t)|^2 = 
\frac{k|C(k)|^2}{\pi^2}\frac{k^4}{a^4}
\left[
1 + \frac{1}{2} f^{abc} u^{b} u^{c} \frac{k|C(k)|^2}{2 \pi^2}
\right]\,,
\label{eq:A-5}
\end{eqnarray} 
where $u^{b}(=1)$ and $u^{c}(=1)$ are the quantities denoting the dependence 
on the indices $b$ and $c$, respectively. 
Thus, from Eq.~(\ref{eq:A-5}) we see that 
the YM magnetic fields evolves 
as $|{B_i^{a}}^{(\mathrm{proper})}(t)|^2 = |\bar{B}^{a}|^2/a^4$, where 
$|\bar{B}^{a}|$ is a constant. 

In this case, we find that
Eqs.~(\ref{eq:2.38}) and (\ref{eq:2.40}) are reduced to 
\begin{eqnarray}
H^2
\Eqn{=}
\kappa^2
\biggl\{
\frac{1}{6} I(R) (\varepsilon - b\tilde{g}^2) +
\left[
-f^{\prime}(R) \left( \dot{H} + H^2 \right) +
6 f^{\prime\prime}(R) H \left( \ddot{H} + 4H\dot{H} \right)
\right] (\varepsilon - b\tilde{g}^2)
\nonumber \\[2mm]
&&
{}-4f^{\prime}(R) H^2 \varepsilon
\biggr\} 
\frac{|\bar{B}^{a}|^2}{a^4}\,,
\label{eq:2.46}
\end{eqnarray}
and
\begin{eqnarray}
2\dot{H} + 3H^2
\Eqn{=} \kappa^2
\biggl\{
\frac{1}{6} I(R) (\varepsilon - 3b\tilde{g}^2) +
\biggl[
-f^{\prime}(R) \left( \dot{H} + 3H^2 \right)
\nonumber \\[2mm]
&& \hspace{-25mm}
{}+6f^{\prime\prime}(R)
\left(\dddot{H}+7H\ddot{H}+4\dot{H}^2+12H^2\dot{H} \right)
+36f^{\prime\prime\prime}(R) \left( \ddot{H}+4H\dot{H} \right)^2
\biggr](\varepsilon - b\tilde{g}^2)
\nonumber \\[2mm]
&& \hspace{-25mm}
{}+4\left[
f^{\prime}(R) \left( -\dot{H} + H^2 \right) -12f^{\prime\prime}(R) H
\left( \ddot{H} + 4H\dot{H}  \right)
\right] \varepsilon + 16 f^{\prime}(R) H^2 b\tilde{g}^2
\biggr\}
\frac{|\bar{B}^{a}|^2}{a^4}\,, 
\label{eq:2.47}
\end{eqnarray}
respectively.
Eliminating $I(R)$ from Eqs.~(\ref{eq:2.46}) and (\ref{eq:2.47}),
we obtain
\begin{eqnarray}
\dot{H} + \frac{\varepsilon}{\varepsilon - b\tilde{g}^2} H^2
\Eqn{=} \kappa^2
\biggl(
f^{\prime}(R) \left\{ -(2 \varepsilon + b\tilde{g}^2)\dot{H}
+ \left[ \left( \frac{3\varepsilon - 7b\tilde{g}^2}
{\varepsilon - b\tilde{g}^2} \right) \varepsilon + 8b\tilde{g}^2
\right] H^2
\right\}
\nonumber \\[2mm]
&&
{}+3f^{\prime\prime}(R)
\left[ (\varepsilon - b\tilde{g}^2)\dddot{H}
- 2(\varepsilon + 2b\tilde{g}^2)H\ddot{H}
+4(\varepsilon - b\tilde{g}^2)\dot{H}^2 - 24 \varepsilon H^2\dot{H}
\right]
\nonumber \\[2mm]
&&
{}+18f^{\prime\prime\prime}(R) (\varepsilon - b\tilde{g}^2)
\left( \ddot{H} + 4H\dot{H} \right)^2
\biggr) 
\frac{|\bar{B}^{a}|^2}{a^4}\,.
\label{eq:2.48}
\end{eqnarray}

We here note the following point. From Eq.~(\ref{eq:2.11}),
we see that the value of $\varepsilon$
depends on the field strength, in other words, it varies in time.
In fact, however, the change in time of $\varepsilon$ is smaller than
that of other quantities
because the dependence of $\varepsilon$ on the field strength is
logarithmic, so that we can approximately regard $\varepsilon$ as
constant in Eq.~(\ref{eq:2.48}).
(Thus, from this point we regard $\varepsilon$ as constant.)

Here we consider the case in which $f(R)$ is given by the following form:
\begin{eqnarray}
f(R) = f_{\mathrm{HS}}(R) \equiv \frac{c_1 \left(R/m^2 \right)^n}
{c_2 \left(R/m^2 \right)^n + 1}\,,
\label{eq:2.49}
\end{eqnarray}
which satisfies the conditions:
\begin{eqnarray}
\lim_{R\to\infty} f_{\mathrm{HS}}(R)
\Eqn{=} \frac{c_1}{c_2} = \mbox{const}\,,
\label{eq:2.50} \\[2mm]
\lim_{R\to 0} f_{\mathrm{HS}}(R)
\Eqn{=} 0\,.
\label{eq:2.51}
\end{eqnarray}
Here, $c_1$ and $c_2$ are dimensionless constants, $n$ is a positive
constant, and $m$ denotes a mass scale. This form, $f_{\mathrm{HS}}(R)$, has
been proposed by Hu and Sawicki~\cite{Hu:2007nk}.
The second condition (\ref{eq:2.51}) means that there could exist a flat
spacetime solution.
Hence, because in the late time universe the value of the scalar curvature
becomes zero, the YM coupling $I$ becomes unity, so
that the standard YM theory can be naturally recovered.

In order to show that power-law inflation can be realized,
we consider the case in which the scale factor is given by
$a(t) = \bar{a} \left(t/\bar{t}\right)^p$,
where $\bar{t}$ is some fiducial time during inflation,
$\bar{a}$ is the value of $a(t)$ at $t=\bar{t}$,
and $p$ is a positive constant.
In this case, $H=p/t$, $\dot{H}=-p/t^2$,
$\ddot{H}=2p/t^3$, and $\dddot{H}=-6p/t^4$. Moreover, it follows from
the fourth equation in (\ref{eq:2.15}) that $R=6p(2p-1)/t^2$.
At the inflationary stage, because $R/m^2 \gg 1$, we are able to
use the following approximate relations:
\begin{eqnarray}
f_{\mathrm{HS}}(R) 
= 
\frac{c_1}{c_2} \left[1-\frac{1}{c_2}
\left( \frac{R}{m^2} \right)^{-n} \right]\,.
\label{eq:2.52} 
\end{eqnarray}
Substituting the above relations in terms of $a$, $H$ and $R$, and
the approximate expressions of 
$f_{\mathrm{HS}}^{\prime}(R)$, 
$f_{\mathrm{HS}}^{\prime\prime}(R)$ and
$f_{\mathrm{HS}}^{\prime\prime\prime}(R)$ derived from
Eq.~(\ref{eq:2.52}) into Eq.~(\ref{eq:2.48}), we find
\begin{eqnarray}
\hspace{-10mm}
p \Eqn{=} 
\frac{n+1}{2}\,,
\label{eq:2.54} 
\\[2mm]
\hspace{-10mm}
\frac{\bar{a}}{\bar{t}^p} 
\Eqn{=}
\biggl\{
\frac{1}{3^{n+1} n^n (n+1)^{n+1}}
\frac{(-c_1)}{{c_2}^2}
\frac{(n+1){\varepsilon}^2 + 3(n-1)b\tilde{g}^2 \varepsilon
+ 6 \left( b\tilde{g}^2 \right)^2}{(n-1)\varepsilon + 2b\tilde{g}^2}
|\bar{B}^{a}|^2 
\kappa^2 m^{2n} \biggr\}^{1/4}\,.
\label{eq:2.55}
\end{eqnarray}
Hence, if $n \gg 1$, $p$ becomes much larger than unity, so that
power-law inflation can be realized.
Consequently, it follows form this result that
the YM field with a non-minimal gravitational
coupling in Eq.~(\ref{eq:2.3}) can be a source of inflation.
This result is the same as in non-minimal Maxwell theory~\cite{Bamba:2008ja}.

In this paper we consider only the case in which the values of
the terms proportional to
$f^{\prime}(R)$, $f^{\prime\prime}(R)$ and $f^{\prime\prime\prime}(R)$
in the right-hand side of
Eqs.~(\ref{eq:2.46}) and (\ref{eq:2.47}) are dominant to the value of
the term proportional to $I(R)$.
Among the terms proportional to
$f^{\prime}(R)$, $f^{\prime\prime}(R)$ and $f^{\prime\prime\prime}(R)$,
the term proportional to $f^{\prime}(R)$ is dominant, and its value is
order $f^{\prime}(R) H^2 \approx n \left(c_1/{c_2}^2\right)
\left( H^2/m^2 \right) \left( R/m^2 \right)^{-(n+1)}$, which can be derived by 
using Eq.~(\ref{eq:2.52}).
Here, it follows from $H=p/t$ and $R=6p(2p-1)/t^2$ that $R$ is order $10 H^2$.
The condition that the term proportional to $f^{\prime}(R)$ is
dominant in the source term would be
$I(R)/\left[ f^{\prime}(R) H^2 \right] \sim 10 c_2 (R/m^2)^n/n \ll 1$.
This would require extremely small $c_2$ because at the inflationary stage
$R/m^2 \gg 1$ and $n \gg 1$. 
In such a case, the value of the right-hand side of Eq.~(\ref{eq:2.48}),
which is order $\kappa^2 f^{\prime}(R) H^2 |\bar{B}^{a}|^2/a^4$, 
can be order $H^2$.
Consequently, the right-hand side of Eq.~(\ref{eq:2.48}) can balance with
the left-hand side of Eq.~(\ref{eq:2.48}), and hence Eq.~(\ref{eq:2.48})
can be satisfied without contradiction to the result, i.e.,
power-law inflation in which $p$ is much larger than unity can be realized.
The reason why we consider the case in which
the term proportional to $I(R)$ on the right-hand side of
Eqs.~(\ref{eq:2.46}) and (\ref{eq:2.47}) is so small in comparison with
the term proportional to $f^{\prime}(R)$ that it can be neglected is
as follows~\cite{Bamba:2008ja}: 
If the opposite case, namely, 
the term proportional to $I(R)$ is dominant to the term proportional to
$f^{\prime}(R)$, Eqs.~(\ref{eq:2.46}) and (\ref{eq:2.47}) are approximately
written as $H^2 \approx (1/6) \kappa^2 I(R) |\bar{B}^{a}|^2/a^4$ and 
$2\dot{H}+3H^2 \approx (1/6) \kappa^2 I(R) |\bar{B}^{a}|^2/a^4$, respectively. 
Thus, in this case it follows from Eqs.~(\ref{eq:2.46}) and (\ref{eq:2.47})
that $H^2$ and $2\dot{H}+3H^2$ are the same order and their difference, 
$2\dot{H}+2H^2$, must be much smaller than $H^2$. In fact, 
Eq.~(\ref{eq:2.48}) implies that 
$\dot{H}+ \left[ \varepsilon/\left(\varepsilon - b\tilde{g}^2\right) \right] 
H^2$ balances with much smaller 
quantity than $\kappa^2 I(R) |\bar{B}^{a}|^2/a^4$. 
Now, 
$\left\{ \dot{H}+ \left[ \varepsilon/\left(\varepsilon - b\tilde{g}^2\right) 
\right] H^2 \right \}/H^2 = 
\varepsilon/\left(\varepsilon - b\tilde{g}^2\right) - 1/p \ll 1$
and hence $p$ must be smaller than unity because $\varepsilon >0 $ and
$b > 0$. Consequently, in this case power-law inflation cannot be realized.

Finally, we note the following point. 
The constraint on a non-minimal gravitational coupling of matter from
the observational data of the central temperature of the Sun has been
proposed~\cite{Bertolami:2007vu}.
Furthermore, the existence of the non-minimal gravitational coupling of
the electromagnetic field changes the
value of the fine structure constant, i.e., the strength of the
electromagnetic coupling. Hence, the deviation of the non-minimal
electromagnetism from the ordinary Maxwell theory can be constrained
from the observations of radio and optical quasar absorption
lines~\cite{Tzanavaris:2006uf}, those of the anisotropy of the cosmic
microwave background (CMB) radiation~\cite{Battye:2000ds, Stefanescu:2007aa},
those of the absorption of CMB radiation at 21 cm hyperfine transition of
the neutral atomic hydrogen~\cite{Khatri:2007yv},
and big bang nucleosynthesis (BBN)~\cite{Bergstrom:1999wm, Avelino:2001nr}
as well as solar-system experiments~\cite{Fujii:2006ic}
(for a recent review, see~\cite{GarciaBerro:2007ir}).
On the other hand, because the energy scale of the YM theory is higher than
the electroweak scale, the existence the non-minimal gravitational coupling of
YM field might influence on models of the grand unified theories (GUT).

\section{Inflation and late-time cosmic acceleration in modified gravity}

Next, in this section we consider a non-minimal gravitational coupling of
the YM field in a modified gravitational theory proposed
in Ref.~\cite{Nojiri:2007as}.

We consider the following model action:
\begin{eqnarray}
S_{\mathrm{MG}}
\Eqn{=}
\int d^{4}x \sqrt{-g}
\left[ \hspace{1mm}
{\mathcal{L}}_{\mathrm{MG}}
+{\mathcal{L}}_{\mathrm{YM}}
\hspace{1mm} \right]\,,
\label{eq:3.1} \\[2mm]
{\mathcal{L}}_{\mathrm{MG}}
\Eqn{=}
\frac{1}{2\kappa^2} \left[ R+F(R) \right]\,,
\label{eq:3.2}
\end{eqnarray}
where $F(R)$ is an arbitrary function of $R$.
Here, ${\mathcal{L}}_{\mathrm{YM}}$ is given by Eq.~(\ref{eq:2.3}).
We note that $F(R)$ is the modified part of gravity, and hence
$F(R)$ is completely different from the non-minimal gravitational
coupling of the YM field $f(R)$ in Eq.\ (\ref{eq:2.4}).

Taking variations of the
action Eq.\ (\ref{eq:3.1}) with respect to the
metric $g_{\mu\nu}$, we find that the field equation of modified gravity
is given by~\cite{Nojiri:2007as}
\begin{eqnarray}
\left[ 1+F^{\prime}(R) \right] R_{\mu \nu}
- \frac{1}{2}g_{\mu \nu} \left[ R+F(R) \right] + g_{\mu \nu}
\Box F^{\prime}(R) - {\nabla}_{\mu} {\nabla}_{\nu} F^{\prime}(R)
= \kappa^2 T^{(\mathrm{YM})}_{\mu \nu}\,.
\label{eq:3.3}
\end{eqnarray}

The $(\mu,\nu)=(0,0)$ component and
the trace part of the $(\mu,\nu)=(i,j)$ component of Eq.~(\ref{eq:3.3}),
where $i$ and $j$ run from $1$ to $3$, are given by
Eqs.~(\ref{eq:2.38}) and (\ref{eq:2.40}), respectively. 

Here we consider the same case as in the preceding section. 
In this case, 
eliminating $I(R)$ from Eqs.~(\ref{eq:2.38}) and (\ref{eq:2.40}), 
we obtain
\begin{eqnarray}
&& \hspace{-15mm}
\dot{H} + \frac{\varepsilon}{\varepsilon - b\tilde{g}^2} H^2
+ \biggl\{ \frac{\varepsilon}{6(\varepsilon - b\tilde{g}^2)} F(R)
- F^{\prime}(R)
\left( \frac{b\tilde{g}^2}{\varepsilon - b\tilde{g}^2} \dot{H} +
\frac{\varepsilon}{\varepsilon - b\tilde{g}^2} H^2 \right)
\nonumber \\[2mm]
&& \hspace{-15mm}
{}+ 3F^{\prime\prime}(R) \left[
\dddot{H} + 4\left( \dot{H}^2 + H\ddot{H} \right) \right]
+ 18F^{\prime\prime\prime}(R) \left( \ddot{H} + 4H\dot{H} \right)^2
\biggr\}
\nonumber 
\end{eqnarray}
\begin{eqnarray}
{}\Eqn{=} \kappa^2
\biggl(
f^{\prime}(R) \left\{ -(2 \varepsilon + b\tilde{g}^2)\dot{H}
+ \left[ \left( \frac{3\varepsilon - 7b\tilde{g}^2}
{\varepsilon - b\tilde{g}^2} \right) \varepsilon
+ 8b\tilde{g}^2 \right] H^2
\right\}
\nonumber \\[2mm]
&&
{}+
3f^{\prime\prime}(R)
\left[ (\varepsilon - b\tilde{g}^2)\dddot{H}
- 2(\varepsilon + 2b\tilde{g}^2)H\ddot{H}
+4(\varepsilon - b\tilde{g}^2)\dot{H}^2 - 24 \varepsilon H^2\dot{H}
\right]
\nonumber \\[2mm]
&&
{}+18f^{\prime\prime\prime}(R) (\varepsilon - b\tilde{g}^2)
\left( \ddot{H} + 4H\dot{H} \right)^2
\biggr)
\frac{|\bar{B}^{a}|^2}{a^4}\,. 
\label{eq:3.4}
\end{eqnarray}

Here we consider the case in which $F(R)$ is given by
\begin{eqnarray}
F(R) \Eqn{=}
- M^2 \frac{\left[ \left(R/M^2\right) - \left(R_0/M^2\right) \right]^{2l+1}
+ {\left(R_0/M^2\right)}^{2l+1}}
{c_3 + c_4 \left\{
\left[ \left(R/M^2\right) - \left(R_0/M^2\right) \right]^{2l+1} +
{\left(R_0/M^2\right)}^{2l+1} \right\}}\,,
\label{eq:3.5}
\end{eqnarray}
which satisfies the following conditions:
$
\lim_{R\to\infty} F(R) = -M^2/c_4 = \mbox{const},
$
$
\lim_{R\to 0} F(R) = 0.
$
Here, $c_3$ and $c_4$ are dimensionless constants,
$l$ is a positive integer, and $M$ denotes a mass scale.
We consider that in the limit $R\to\infty$, i.e., at the very early stage of
the universe, $F(R)$ becomes an effective cosmological constant, 
$
\lim_{R\to\infty} F(R) = -M^2/c_4 = -2{\Lambda}_{\mathrm{i}}, 
$ 
where 
${\Lambda}_{\mathrm{i}} \left(\gg {H_0}^2 \right)$ is an effective
cosmological constant in the very early universe, 
and that
at the present time $F(R)$ becomes a small constant, 
$
F(R_0) = -M^2 \left(R_0/M^2\right)^{2l+1}/ 
\left[ {c_3 + c_4 \left(R_0/M^2\right)^{2l+1}} \right] = -2R_0, 
$
where $R_0 \left(\approx {H_0}^2 \right)$ is current curvature. 
Here, 
$H_0$ is the Hubble constant at the present time:
$H_{0} = 100 h \hspace{1mm} \mathrm{km} \hspace{1mm} {\mathrm{s}}^{-1}
\hspace{1mm} {\mathrm{Mpc}}^{-1}
= 2.1 h \times 10^{-42} {\mathrm{GeV}}
\approx 1.5 \times 10^{-33} {\mathrm{eV}}$~\cite{Kolb and Turner}, where
we have used $h=0.70$~\cite{Freedman:2000cf}. 

Furthermore, we consider the case in which $f(R)$ is given by the
following form:
\begin{eqnarray}
f(R) = f_{\mathrm{NO}}(R) \equiv
\frac{\left[ \left(R/M^2\right) - \left(R_0/M^2\right) \right]^{2q+1}
+ {\left(R_0/M^2\right)}^{2q+1}}
{c_5 + c_6 \left\{
\left[ \left(R/M^2\right) - \left(R_0/M^2\right) \right]^{2q+1} +
{\left(R_0/M^2\right)}^{2q+1} \right\}}\,,
\label{eq:3.10}
\end{eqnarray}
which satisfies the following conditions:
$
\lim_{R\to\infty} f_{\mathrm{NO}}(R) = 1/c_6 = \mbox{const},
$
$
\lim_{R\to 0} f_{\mathrm{NO}}(R) = 0.
$
Here, $c_5$ and $c_6$ are dimensionless constants, and
$q$ is a positive integer.
The form of $F(R)$ in Eq.~(\ref{eq:3.5}) and
$f_{\mathrm{NO}}(R)$ in Eq.~(\ref{eq:3.10})
is taken from Ref.~\cite{Nojiri:2007as}. This
form corresponds to the extension of the form of $f_{\mathrm{HS}}(R)$ in
Eq.~(\ref{eq:2.49}). It has been shown in Ref.~\cite{Nojiri:2007as}
that modified gravitational theories described by
the action (\ref{eq:3.2}) with $F(R)$ in Eq.~(\ref{eq:3.5}) successfully
pass the solar-system tests as well as cosmological bounds
and they are free of instabilities.

Making the same considerations as in Ref.~\cite{Bamba:2008ja}, we 
find that 
at the very early stage of the universe, it follows from Eq.~(\ref{eq:3.4}) 
that 
$
a(t) \propto \exp \left(\sqrt{{\Lambda}_{\mathrm{i}}/3} t \right), 
$ 
so that exponential inflation can be realized, 
and that at the present time, it follows from Eq.~(\ref{eq:3.4}) that 
$ 
a(t) \propto \exp \left(\sqrt{R_0/3} t \right), 
$ 
so that the late-time acceleration of the universe can be realized. 
These results are also the same as in non-minimal Maxwell-$F(R)$
gravity~\cite{Bamba:2008ja}. 

Finally, we note the following point about the logarithmic contribution
to modified gravity, namely, the case in which
the Lagrangian of modified gravity in Eq.~(\ref{eq:3.2}) are given by
${\mathcal{L}}_{\mathrm{MG}} =
1/\left(2\kappa^2\right) \left[ R+F(R)+\ln\left(R/M^2 \right) \right]$.
Following to the considerations in the previous subsections,
because the logarithmic term is sub-leading contribution,
in also this case both inflation and the late-time acceleration of the universe can be realized. The qualitative difference from the case of
the previous subsections is only that in the limit $R\to\infty$
the gravitational modification term,
$F(R) + \ln\left(R/M^2 \right)$ does not become constant. In fact, however,
if it is considered that some cut off scale of $R$ in the very early universe
should exist, the logarithmic contribution does not diverge in this limit, and
hence the cosmology of this case is the same as that of the previous sections.

\section{Non-minimal vector model}

In this section, we consider the cosmology in the non-abelian non-minimal
vector-$F(R)$ gravity.

We consider the following model action:
\begin{eqnarray}
\bar{S}_{\mathrm{MG}}
\Eqn{=}
\int d^{4}x \sqrt{-g}
\left[ \hspace{1mm}
{\mathcal{L}}_{\mathrm{MG}}
+{\mathcal{L}}_{\mathrm{V}}
\hspace{1mm} \right]\,,
\label{eq:4.1} \\[2mm]
{\mathcal{L}}_{\mathrm{V}}
\Eqn{=}
I(R) \left\{
-\frac{1}{4} F_{\mu\nu}^{a} F^{a\mu\nu} - V[A^{a2}]
\right\}\,,
\label{eq:4.2}
\end{eqnarray}
where
${\mathcal{L}}_{\mathrm{MG}}$ is given by Eq.~(\ref{eq:3.2}),
$F_{\mu\nu}^{a}$ is given by Eq.~(\ref{eq:2.6}), and
$A^{a2}= g^{\mu \nu} A_{\mu}^{a} A_{\nu}^{a}$.
(As the generalization of the above non-minimal vector model,
one can consider a model in which the derivative in
$F_{\mu\nu}^{a}$ is the gauge covariant derivative given by
$D_\mu = \nabla_\mu -i \tilde{g} A_\mu$, where $A_\mu = A^a_\mu \tau^a$.
Here, $\tau^a$ are matrices and their commutation relations is
conventionally written as the standard form
$\left[ \tau^a, \tau^b \right] = i f^{abc} \tau^c$~\cite{P-S}.
In the present paper, however, as a simple non-minimal vector model
we consider the theory described by the action in Eq.~(\ref{eq:4.2})).

We should note that the last term $V[A^{a2}]$ in the action (\ref{eq:4.2}) is 
not gauge invariant but can be rewritten in a gauge invariant way. 
For example if the gauge group is a unitary group, we may introduce a $\sigma$-model like field $U$, 
which satisfies $U^\dagger U=1$. Then the last term could be rewritten in the
gauge invariant form:
\be
\label{sn1}
V[A^{a2}] \to V\left[ \bar{c} \tr \left(U^\dagger A_{\mu}^{a} U\right)
\left(U^\dagger A^{a\mu} U\right) \right]\ .
\ee
Here $\bar{c}$ is a constant for the normalization.
If we choose the unitary gauge $U=1$, the term in (\ref{sn1}) reduces to
the original one: $V[A^{a2}]$. This may tells that the action (\ref{eq:4.2}) described
the theory where the gauge group is spontaneously broken.

The field equations can be derived by taking variations of the
action in Eq.~(\ref{eq:4.1}) with respect to the
metric $g_{\mu\nu}$ and the vector field $A_{\mu}^{a}$ as follows:
\begin{eqnarray}
\left[ 1+F^{\prime}(R) \right] R_{\mu \nu}
- \frac{1}{2}g_{\mu \nu} \left[ R+F(R) \right] + g_{\mu \nu}
\Box F^{\prime}(R) - {\nabla}_{\mu} {\nabla}_{\nu} F^{\prime}(R)
= \kappa^2 T^{(\mathrm{V})}_{\mu \nu}\,,
\label{eq:4.3}
\end{eqnarray}
with
\begin{eqnarray}
\hspace{-5mm}
T^{(\mathrm{V})}_{\mu \nu}
\Eqn{=}
I(R) \left\{ g^{\alpha\beta} F_{\mu\beta}^{a} F_{\nu\alpha}^{a}
+ 2 A_{\mu}^{a} A_{\nu}^{a}
\frac{d V[A^{a2}]}{d A^{a2}}
-\frac{1}{4} g_{\mu\nu} \bar{\mathcal{F}} \right\}
\nonumber \\[2mm]
&&
{}+\frac{1}{2} \left\{ f^{\prime}(R) \bar{\mathcal{F}} R_{\mu \nu}
+ g_{\mu \nu} \Box \left[ f^{\prime}(R) \bar{\mathcal{F}} \right]
- {\nabla}_{\mu} {\nabla}_{\nu}
\left[ f^{\prime}(R) \bar{\mathcal{F}} \right]
\right\}\,,
\label{eq:4.4} \\[2mm] 
\bar{\mathcal{F}} 
\Eqn{=} 
F_{\mu\nu}^{a} F^{a\mu\nu} + 4V[A^{a2}]\,,
\label{eq:4.5}
\end{eqnarray}
and
\begin{eqnarray}
\frac{1}{\sqrt{-g}}{\partial}_{\mu}
\left[ \sqrt{-g} I(R) F^{a\mu\nu} \right]
- I(R) \left\{
f^{abc} A_{\mu}^{b} F^{c\mu\nu}
+ 2 \frac{d V[A^{a2}]}{d A^{a2}} A^{a \nu}
\right\}
= 0\,,
\label{eq:4.6}
\end{eqnarray}
where $T^{(\mathrm{V})}_{\mu \nu}$ is the contribution to
the energy-momentum tensor from $A_{\mu}^{a}$.

Here, as an example, we consider the case in which $V[A^{a2}]$ is given by
a class of the following power-law potential:
\begin{eqnarray}
V[A^{a2}]=\bar{V} \left( \frac{A^{a2}}{\bar{m}^2} \right)^{\bar{n}}\,,
\label{eq:4.7}
\end{eqnarray}
where $\bar{V}$ is a constant, $\bar{m}$ denotes a mass scale, and
$\bar{n} (>1)$ is a positive integer.

Similarly to Sec.\ II B, 
because the amplitude of $A_{\mu}^{a}$ is small, we neglect 
the higher order than or equal to the quadratic terms in $A_{\mu}^{a}$ 
and consider the linearized equation of Eq.\ (\ref{eq:4.6}) 
in terms of $A_{\mu}^{a}$. 
For the power-law potential given by Eq.\ (\ref{eq:4.7}),
the linearized equation of motion 
under the ansatz 
${\partial}^jA_j^{a}(t,\Vec{x}) =0$ and $A_{0}^{a}(t,\Vec{x}) = 0$ 
is the same as Eq.\ (\ref{eq:A-1}).
\footnote{
This is similar to the Coulomb gauge but since the action (\ref{eq:4.2}) is
not gauge invariant, or gauge symmetry is completely fixed by the unitary gauge as in
after (\ref{sn1}), this condition is only a working hypothesis.
}

The $(\mu,\nu)=(0,0)$ component and
the trace part of the $(\mu,\nu)=(i,j)$ component of Eq.~(\ref{eq:4.3}),
where $i$ and $j$ run from $1$ to $3$, read
\begin{eqnarray}
&&\hspace{-5mm}
H^2 + \frac{1}{6} F(R)  - F^{\prime}(R) \left( \dot{H} + H^2 \right)
\nonumber \\[2mm]
&&
=
\frac{\kappa^2}{6}
\biggl(
I(R) \left\{ Y + 2V[A^{a2}] \right\}
+ 3 \left[
-f^{\prime}(R) \left( \dot{H} + H^2 \right) +
6 f^{\prime\prime}(R) H \left( \ddot{H} + 4H\dot{H} \right)
\right] \bar{\mathcal{F}}
\nonumber \\[2mm]
&& \hspace{5mm}
{}+ 3 f^{\prime}(R) H \dot{\bar{\mathcal{F}}}
\biggr)\,,
\label{eq:4.8}
\end{eqnarray}
and
\begin{eqnarray}
&& \hspace{-15mm}
2\dot{H} + 3H^2 +\frac{1}{2}F(R)
- F^{\prime}(R) \left( \dot{H} + 3H^2 \right)
\nonumber \\[2mm]
&& \hspace{-15mm}
{}+ 6F^{\prime\prime}(R) \left[
\dddot{H} + 4\left( \dot{H}^2 + H\ddot{H} \right)
\right]
+ 36F^{\prime\prime\prime}(R) \left( \ddot{H} + 4H\dot{H} \right)^2
\nonumber \\[2mm]
{}\Eqn{=}
\frac{\kappa^2}{2}
\biggl(
\frac{1}{3} I(R) \left\{ -X + 6V[A^{a2}]
-4\frac{1}{a^2} A_{i}^{a} A_{i}^{a} \frac{d V[A^{a2}]}{d A^{a2}} \right\}
+ \biggl[
-f^{\prime}(R) \left( \dot{H} + 3H^2 \right)
\nonumber \\[2mm]
&&
{}+6f^{\prime\prime}(R)
\left(\dddot{H}+7H\ddot{H}+4\dot{H}^2+12H^2\dot{H} \right)
+36f^{\prime\prime\prime}(R) \left( \ddot{H}+4H\dot{H} \right)^2
\biggr] \bar{\mathcal{F}}
\nonumber \\[2mm]
&&
{}+3\left[f^{\prime}(R)H + 4f^{\prime\prime}(R)
\left( \ddot{H}+4H\dot{H} \right) \right] \dot{\bar{\mathcal{F}}}
+ f^{\prime}(R) \ddot{\bar{\mathcal{F}}}
\biggr)\,,
\label{eq:4.9}
\end{eqnarray}
respectively.
In deriving Eqs.~(\ref{eq:4.8}) and (\ref{eq:4.9}), we have used
equations in (\ref{eq:2.15}).

Here we consider the same case as in the previous sections. 
Moreover, we here consider the case in which $A_{0}^{a} = 0$. 
In this case, we have 
$\left( 1/a^2 \right) A_{i}^{a} A_{i}^{a} d V[A^{a2}]/ \left( d A^{a2} \right)
= \bar{n} V[A^{a2}]$. 
Consequently, using this relation, we find that
Eqs.~(\ref{eq:4.8}) and (\ref{eq:4.9}) are reduced to
\begin{eqnarray}
&& \hspace{-18mm}
H^2 + \frac{1}{6} F(R) - F^{\prime}(R) \left( \dot{H} + H^2 \right)
\nonumber \\[2mm]
&& \hspace{-15mm}
=
\kappa^2 \biggl(
\left[
\frac{1}{6} I(R)
-f^{\prime}(R) \left( \dot{H} + 5H^2 \right) +
6 f^{\prime\prime}(R) H \left( \ddot{H} + 4H\dot{H} \right)
\right]
\frac{|\bar{B}^{a}|^2}{a^4} 
\nonumber \\[2mm]
&& \hspace{-10mm}
{}+
\left\{
\frac{1}{3} I(R) -2f^{\prime}(R)
\left[ \dot{H} + \left(1+2 \bar{n} \right) H^2  \right]
+12f^{\prime\prime}(R) H \left( \ddot{H} + 4H\dot{H} \right)
\right\} V[A^{a2}]
\biggr)\,,
\label{eq:4.11}
\end{eqnarray}
and
\begin{eqnarray}
&& \hspace{-15mm}
2\dot{H} + 3H^2 +\frac{1}{2}F(R)
- F^{\prime}(R) \left( \dot{H} + 3H^2 \right)
\nonumber \\[2mm]
&& \hspace{-15mm}
{}+ 6F^{\prime\prime}(R) \left[
\dddot{H} + 4\left( \dot{H}^2 + H\ddot{H} \right)
\right]
+ 36F^{\prime\prime\prime}(R) \left( \ddot{H} + 4H\dot{H} \right)^2
\nonumber \\[2mm]
\Eqn{=}
\kappa^2 \biggl(
\biggl[
\frac{1}{6} I(R)
+f^{\prime}(R) \left( - 5\dot{H} + H^2 \right) +
6 f^{\prime\prime}(R) \left( \dddot{H} -H\ddot{H}+4\dot{H}^2-20H^2\dot{H}
\right)
\nonumber \\[2mm]
&& \hspace{0mm}
{}+36f^{\prime\prime\prime}(R)
\left( \ddot{H} + 4H\dot{H} \right)^2
\biggr]
\frac{|\bar{B}^{a}|^2}{a^4} 
\nonumber \\[2mm]
&& \hspace{0mm}
{}
+\biggl\{
\frac{1}{3} I(R)
-2f^{\prime}(R) \left[ \left(1+2 \bar{n} \right) \dot{H} +
\left( 3 + 6 \bar{n} - 4\bar{n}^2 \right) H^2
\right]
\nonumber \\[2mm]
&& \hspace{0mm}
{}+ 12f^{\prime\prime}(R) \left[
\dddot{H} + \left( 7-4\bar{n} \right) H\ddot{H} + 4\dot{H}^2
+ 4\left( 3-4\bar{n} \right) H^2\dot{H}
\right]
\nonumber \\[2mm]
&& \hspace{0mm}
{}+ 72f^{\prime\prime\prime}(R) \left( \ddot{H} + 4H\dot{H} \right)^2
\biggr\} V[A^{a2}]
\biggr)\,,
\label{eq:4.12}
\end{eqnarray}
respectively.
Eliminating $I(R)$ from Eqs.~(\ref{eq:4.11}) and (\ref{eq:4.12}),
we obtain
\begin{eqnarray}
&& \hspace{-4.5mm}
\dot{H} + H^2 +
\left\{ \frac{1}{6} F(R) - F^{\prime}(R) H^2 +
3F^{\prime\prime}(R) \left[
\dddot{H} + 4\left( \dot{H}^2 + H\ddot{H} \right) \right]
+ 18F^{\prime\prime\prime}(R) \left( \ddot{H} + 4H\dot{H} \right)^2
\right\}
\nonumber \\[2mm]
&& \hspace{5mm}
{}=
\kappa^2 \biggl(
\biggl[
f^{\prime}(R) \left( -2\dot{H} + 3H^2 \right) +
3f^{\prime\prime}(R) \left( \dddot{H}-2H\ddot{H}+4\dot{H}^2-24H^2\dot{H}
\right)
\nonumber \\[2mm]
&& \hspace{10mm}
{}+18f^{\prime\prime\prime}(R)
\left( \ddot{H} + 4H\dot{H} \right)^2
\biggr]
\frac{|\bar{B}^{a}|^2}{a^4} 
+2 \biggl\{
-f^{\prime}(R) \left[ \bar{n} \dot{H} +
\left( 1 + 2 \bar{n} - 2\bar{n}^2 \right) H^2
\right]
\nonumber \\[2mm]
&& \hspace{10mm}
{}+3f^{\prime\prime}(R) \left[
\dddot{H} + 2\left( 3-2\bar{n} \right) H\ddot{H} + 4\dot{H}^2
+ 8\left( 1-2\bar{n} \right) H^2\dot{H}
\right]
\nonumber 
\end{eqnarray}
\begin{eqnarray}
{}+18f^{\prime\prime\prime}(R) \left( \ddot{H} + 4H\dot{H} \right)^2
\biggr\} V[A^{a2}]
\biggr)\,.
\label{eq:4.13}
\end{eqnarray}

In the case that $|{B_i^{a}}^{(\mathrm{proper})}(t)|^2 = |\bar{B}^{a}|^2/a^4$, 
$V[A^{a2}] \propto a^{-2\bar{n}}$. Hence, if $\bar{n}=2$, 
the time evolution of $V[A^{a2}]$ is the same as that of 
$|{B_i^{a}}^{(\mathrm{proper})}(t)|^2$. On the other hand, if
$\bar{n} \geq 2$, $V[A^{a2}]$ decreases much more rapidly than 
$|{B_i^{a}}^{(\mathrm{proper})}(t)|^2$ during inflation. Thus, 
in the latter case we can neglect the terms proportional to $V[A^{a2}]$
on the right-hand side of Eq.~(\ref{eq:4.13}).
Consequently, it follows from Eq.~(\ref{eq:4.13}) that
when we consider the case in which similarly to the preceding
section, $F(R)$ and $f(R)$ are
given by Eqs.~(\ref{eq:3.5}) and (\ref{eq:3.10}), respectively,
we can make the same consideration as the preceding section, and
hence power-law inflation and the late-time acceleration of the universe
can be realized.

Furthermore, as another case, we consider the case in which
$F(R)$ is given by Eq.~(\ref{eq:3.5}) and $f(R)$ is given by the following
form:
\begin{eqnarray}
f(R) = \bar{f}(R) \equiv \frac{c_7 \left(R/\bar{M}^2 \right)^{\bar{q}}-1}
{c_8 \left(R/\bar{M}^2 \right)^{\bar{q}} + 1}\,,
\label{eq:4.14}
\end{eqnarray}
which satisfies the following conditions:
$
\lim_{R\to\infty} \bar{f}(R) = c_7/c_8 = \mbox{const},
$
$
\lim_{R\to 0} \bar{f}(R) = -1.
$
Here, $c_7$ and $c_8$ are dimensionless constants, $\bar{q}$ is a positive
constant, and $\bar{M}$ denotes a mass scale.
In this case, the value of
$I(R) = 1 + f(R)$ becomes close to zero when that of $R$ is very small,
namely, at the present time.
Making the same consideration as the preceding section, we can 
also find in this case that power-law inflation and the late-time acceleration
of the universe can be realized.

\section{Duality}

In this section, we consider the duality of the non-minimal
electromagnetic theory and that of the non-minimal YM theory.

\subsection{Duality of the non-minimal electromagnetic theory}

We consider the duality of the action of $\tilde{f}(R)$-coupled electromagnetic theory:
\be
S_{\tilde{f}A} = \frac{1}{4} \int d^4 x \sqrt{-g} \tilde{f}(R)
F_{A\,\mu\nu} F_A^{\mu\nu}\ ,\quad
F_{A\, \mu\nu} \equiv \partial_\mu A_\nu - \partial_\nu A_\mu\ ,
\label{eq:5.1}
\ee
where $\tilde{f}(R)$ is an arbitrary function of $R$ and
$A_\nu$ is the $U(1)$ gauge field.
Before going to the duality of the action in Eq.~(\ref{eq:5.1}), we consider
the duality without gravity:
\be
S_{A} = \frac{1}{4} \int d^4 x F_{A\,\mu\nu} F_{\,A}^{\mu\nu}\ ,\quad
F_{A\, \mu\nu} \equiv \partial_\mu A_\nu - \partial_\nu A_\mu \ .
\label{eq:5.2}
\ee
By introducing a new field $\bar{B}_\mu$, the action can be rewritten as
\be
S_{\bar{F}\bar{B}} =
\frac{1}{4} \int d^4 x \left( \bar{F}_{\,\mu\nu} \bar{F}^{\mu\nu}
+ \frac{1}{2} \epsilon^{\mu\nu\rho\sigma} \left(\partial_\mu \bar{B}_\nu\right)
\bar{F}_{\rho\sigma} \right) \ .
\label{eq:5.3}
\ee
Here, $\bar{F}_{\mu\nu}$ is an independent field
(not given in terms of $A_\mu$ or $\bar{B}_\mu$ as in $F_{A\,\mu\nu}$).
The variation of $\bar{B}_\mu$ gives
\be
\epsilon^{\mu\nu\rho\sigma} {\partial}_\nu \bar{F}_{\rho\sigma} = 0\ ,
\label{eq:5.4}
\ee
which tells that $\bar{F}_{\mu\nu}$ can be given in terms of
a vector field $A_\mu$ as
\be
\bar{F}_{\mu\nu} = F_{A\, \mu\nu} = \partial_\mu A_\nu - \partial_\nu A_\mu \ .
\label{eq:5.5}
\ee
Then the action $S_{\bar{F}\bar{B}}$ in Eq.~(\ref{eq:5.3}) reduces to
$S_A$ in Eq.~(\ref{eq:5.2}).
On the other hand, by the variation of $\bar{F}_{\mu\nu}$, we obtain
\be
\bar{F}^{\mu\nu} = - \frac{1}{4} \epsilon^{\mu\nu\rho\sigma}
\partial_\rho \bar{B}_\sigma\ .
\label{eq:5.6}
\ee
By substituting Eq.~(\ref{eq:5.6}) into the action in Eq.~(\ref{eq:5.2}),
we obtain
\be
S_{\bar{B}} = \frac{1}{4} \int d^4 x F_{\bar{B}\,\mu\nu}
F_{\,\bar{B}}^{\mu\nu}\ ,\quad
F_{\bar{B}\, \mu\nu} \equiv \partial_\mu \bar{B}_\nu - \partial_\nu \bar{B}_\mu \ .
\label{eq:5.7}
\ee
Eqs.~(\ref{eq:5.5}) and (\ref{eq:5.6}) give
\be
F_A^{\mu\nu}=\frac{1}{2} \epsilon^{\mu\nu\rho\sigma} F_{\bar{B}\,\rho\sigma}\ ,
\label{eq:5.8}
\ee
which tells that $F_{B\, \mu\nu}$ are dual to $F_{A\, \mu\nu}$, that is, the magnetic field exchanges
with the electric field.

We now consider the action in Eq.~(\ref{eq:5.1}), which can be rewritten as
\be
S_{\tilde{f}\bar{F}\bar{B}} =
\frac{1}{4} \int d^4 x \left\{ \sqrt{-g} \tilde{f}(R)
\bar{F}_{\mu\nu} \bar{F}^{\mu\nu}
+ \frac{1}{2} \epsilon^{\mu\nu\rho\sigma}
\left(\partial_\mu \bar{B}_\nu\right) \bar{F}_{\rho\sigma} \right\}\ .
\label{eq:5.9}
\ee
Now $\bar{F}_{\mu\nu}$ is an independent field again. From the variation
of $\bar{B}_\mu$, we obtain Eq.~(\ref{eq:5.4}),
which can be solved as Eq.~(\ref{eq:5.5}),
and we find the action $S_{\tilde{f}\bar{F}\bar{B}}$
in Eq.~(\ref{eq:5.9}) is equivalent to Eq.~(\ref{eq:5.1}).
On the other hand, by the variation of $\bar{F}_{\mu\nu}$,
instead of Eq.~(\ref{eq:5.6}), we obtain
\be
\bar{F}^{\mu\nu} = - \frac{1}{4} \frac{ \epsilon^{\mu\nu\rho\sigma}
\partial_\rho \bar{B}_\sigma}{\tilde{f}(R)\sqrt{-g}}\ .
\label{eq:5.10}
\ee
Then substituting Eq.~(\ref{eq:5.10}) into Eq.~(\ref{eq:5.9}) and
using the identity
\be
\epsilon^{\alpha\beta\rho\sigma} \epsilon^{\mu\nu\gamma\delta} g_{\mu\alpha} g_{\nu\beta}
= 2 g \left(g^{\rho\gamma} g^{\sigma\delta} -
g^{\rho \delta} g^{\sigma\gamma} \right)\ ,
\label{eq:5.11}
\ee
we obtain an action dual to Eq.~(\ref{eq:5.1}):
\be
S_{fB} = \frac{1}{4} \int d^4 x \sqrt{-g} \frac{1}{\tilde{f}(R)}
F_{\bar{B}\,\mu\nu} F_{\bar{B}}^{\mu\nu}\ ,\quad
F_{\bar{B}\, \mu\nu} \equiv \partial_\mu \bar{B}_\nu - \partial_\nu \bar{B}_\mu \ .
\label{eq:5.12}
\ee

\subsection{Duality of the non-minimal Yang-Mills theory}

As in case of the electromagnetic theory, we may consider the duality of the action of
$\tilde{f}(R)$-coupled Yang-Mills theory:
\be
S_{\tilde{f}A} = \frac{1}{4} \int d^4 x \sqrt{-g} \tilde{f}(R)
F_{A\,\mu\nu} F_A^{\mu\nu}\ ,\quad
F_{A\, \mu\nu} \equiv \partial_\mu A_\nu - \partial_\nu A_\mu + f^{abc} A^b_\mu A^c_\nu\ .
\label{eq:5.1b}
\ee
The action can be rewritten in the following form
\be
S_{\tilde{f}\bar{F}\bar{B}} =
\frac{1}{4} \int d^4 x \left\{ \sqrt{-g} \tilde{f}(R)
\bar{F}_{\mu\nu} \bar{F}^{\mu\nu}
+ \frac{1}{4} \epsilon^{\mu\nu\rho\sigma}
F^A_{\mu\nu} \bar{F}_{\rho\sigma} \right\}\ .
\label{eq:5.9b}
\ee
Now $\bar{F}_{\mu\nu}$ is an independent field again. From the variation
of $A^a_\mu$ in $F^a_{\mu\nu}$, we obtain
\be
\epsilon^{\mu\nu\rho\sigma}
D_\nu \bar{F}_{\rho\sigma} =0 \ .
\label{eq:5.4b}
\ee
Here $D_\mu$ is a covariant derivative.
The solution of (\ref{eq:5.4b}) is given by
\be
\bar{F}_{\mu\nu}=F_{A\,\mu\nu}\ .
\label{eq:5.5b}
\ee
By substituting (\ref{eq:5.5b}) into (\ref{eq:5.9b}), we obtain (\ref{eq:5.1b}).
On the other hand, by the variation of $\bar{F}_{\mu\nu}$, we obtain
\be
\bar{F}^{\mu\nu} = - \frac{1}{8} \frac{ \epsilon^{\mu\nu\rho\sigma}
F_{A\,\rho\sigma}}{\tilde{f}(R)\sqrt{-g}}\ .
\label{eq:5.10b}
\ee
Then substituting Eq.~(\ref{eq:5.10b}) into Eq.~(\ref{eq:5.9b}) and
using the identity (\ref{eq:5.11}),
we obtain a dual action:
\be
S_{fB} = \frac{1}{4} \int d^4 x \sqrt{-g} \frac{1}{\tilde{f}(R)}
F_{A\,\mu\nu} F_{A}^{\mu\nu} \ .
\label{eq:5.12b}
\ee
Note that dual form of the action maybe useful in the cosmological
considerations.

\section{Reconstruction of the YM theory}
In this section, we indicate how to reconstruct the YM theory from the 
known universe evolution (for a review, see~\cite{reconstruction}).

We now consider the following action:
\be
\label{YM1}
S=\int d^4 x \sqrt{-g}\left(\frac{R}{2\kappa^2} + \tilde{\cal F}
\left(F^a_{\mu\nu}F^{a\,\mu\nu}\right)
\right)\ .
\ee
By introducing an auxiliary scalar field $\phi$, we may rewrite the action
(\ref{YM1}) in the
following form:
\be
\label{YM2}
S=\int d^4 x \sqrt{-g}\left(\frac{R}{2\kappa^2} + \frac{1}{4}P(\phi)
F^a_{\mu\nu}F^{a\,\mu\nu}
+ \frac{1}{4}Q(\phi) \right)\ .
\ee
By the variation of $\phi$, we obtain
\be
\label{YM3}
0= P'(\phi) F^a_{\mu\nu}F^{a\,\mu\nu} + Q'(\phi)\ ,
\ee
which could be solved with respect $\phi$ as
$\phi=\phi\left(F^a_{\mu\nu}F^{a\,\mu\nu}\right)$.
Here, the prime denotes differentiation with respect to $\phi$.
Then by substituting the expression into (\ref{YM2}) we obtain the action
(\ref{YM1}) with
\be
\label{YM4}
\tilde{\cal F}\left(F^a_{\mu\nu}F^{a\,\mu\nu}\right) = \frac{1}{4}\left\{
P\left(\phi\left(F^a_{\mu\nu}F^{a\,\mu\nu}\right)\right)
F^a_{\mu\nu}F^{a\,\mu\nu}
+ Q\left(\phi\left(F^a_{\mu\nu}F^{a\,\mu\nu}\right)\right) \right\}\ .
\ee
By the variation of the action (\ref{YM2}) with respect to the metric
tensor $g_{\mu\nu}$,
we obtain the Einstein equation:
\be
\label{YM5}
\frac{1}{2\kappa^2}\left( R_{\mu\nu} - \frac{1}{2}R g_{\mu\nu} \right)
= - \frac{1}{2} P(\phi) F^a_{\mu\rho} F^{a\ \rho}_{\ \nu}
+ \frac{1}{8}g_{\mu\nu} \left( P(\phi) F^a_{\rho\sigma}F^{a\,\rho\sigma} +
Q(\phi) \right)\ .
\ee
On the other hand, by the variation with respect to $A^a_\mu$, we obtain
\be
\label{YM6}
0=\partial_\nu \left( \sqrt{-g}P(\phi)F^{a\,\nu\mu}\right)
 - \sqrt{-g} P(\phi) f^{abc} A^b_\nu F^{c\,\nu\mu}\ .
\ee
For simplicity, we only consider the case that the gauge algebra is $SU(2)$,
where
$f^{abc}=\epsilon^{abc}$, and we assume the gauge
fields are given in the following form
\be
\label{YM7}
A^a_\mu = \left\{ \begin{array}{cl}
\bar{\alpha}\e^{\lambda(t)}\delta^a_{\ i} & (\mu=i=1,2,3) \\
0 & (\mu=0)
\end{array} \right. \ .
\ee
Here $\bar{\alpha}$ is a constant with mass dimension and
$\lambda$ is a proper function of $t$.
In general, if the vector field is condensed, the rotational invariance
of the universe could be broken.
In case of (\ref{YM7}), the direction of the vector field is gauge variant.
Then all the gauge
invariant quantities given by (\ref{YM7}) do not break the rotational
invariance.

By the assumption, (\ref{YM3}) has the following form:
\be
\label{YM8}
0= 6 \left( - \bar{\alpha}^2 {\dot \lambda}^2 \e^{2\lambda} a^{-2} +
\bar{\alpha}^4 \e^{4\lambda} a^{-4} \right)
P'(\phi) + Q'(\phi)\ ,
\ee
and $(t,t)$-component of Eq.~(\ref{YM5}) is given by
\be
\label{YM9}
0=\frac{3}{\kappa^2}H^2 - \frac{3}{2}\left( \bar{\alpha}^2
{\dot \lambda}^2 \e^{2\lambda} a^{-2}
+ \bar{\alpha}^4 \e^{4\lambda} a^{-4} \right) - \frac{1}{4}Q(\phi)\ .
\ee
The $\mu=0$ component of (\ref{YM6}) becomes identity and $\mu=i$ component
gives
\be
\label{YM10}
0= \partial_t \left(a P(\phi) \dot \lambda \e^{\lambda} \right)
 - 2 \bar{\alpha}^2 a^{-1} P(\phi) \e^{3\lambda}\ .
\ee

Since we can always the scalar field $\phi$ properly, we may identify the
scalar field with the
time coordinate $\phi=t$.
Then by differentiating Eq.~(\ref{YM9}) with respect to $t$ and eliminating
$\dot Q= Q'(\phi)$,
we obtain
\bea
\label{YM11}
0 &=& \frac{2}{\kappa^2}H\dot H + \bar{\alpha}^2 {\dot\lambda}^2 \e^{2\lambda}
a^{-2} \dot P \nn
&& - P\left\{ \bar{\alpha}^2
\left(\dot\lambda \ddot\lambda + {\dot\lambda}^3 - \lambda^2 H\right)
\e^{2\lambda} a^{-2} + 2 \bar{\alpha}^4 \left( \dot\lambda - H \right)
\e^{4\lambda} a^{-4} \right\}\ .
\eea
Furthermore by eliminating $\dot P$ by using (\ref{YM10}), we find
\be
\label{YM12}
P=\frac{2H\dot H}{\kappa^2 \left\{ 2\bar{\alpha}^2 a^{-2} \e^{2\lambda}
\left({\dot\lambda}^2
+ \dot\lambda \ddot\lambda \right) - \bar{\alpha}^4 \e^{4\lambda} a^{-4} H
\right\} }\ .
\ee
Then by using (\ref{YM12}), we can eliminate $P$ (and $\dot P$) in
(\ref{YM10}) and
obtain
\bea
\label{YM13}
0 &=& 2 \left(\dot\lambda \dddot\lambda + {\ddot\lambda}^2
+ 3{\dot\lambda}^2 \ddot\lambda \right) - \bar{\alpha}^2
\e^{2\lambda} a^{-2} \dot H
+ 4\left\{{\dot\lambda}^3 + \dot\lambda \ddot\lambda
 - \bar{\alpha}^2
\e^{2\lambda} a^{-2} H \right\}\left(\dot\lambda - H\right) \nn
&& + \left\{2 \left({\dot\lambda}^3 + \dot\lambda \ddot\lambda \right)
- \bar{\alpha}^2 \e^{2\lambda} a^{-2} H \right\} \left\{\frac{\dot H}{H} +
\frac{\ddot H}{\dot H}
+ H + \frac{\ddot\lambda}{\dot\lambda} + \dot\lambda
 - \frac{2\bar{\alpha}^2 a^{-2}\e^{2\lambda}}{\dot\lambda} \right\}\ .
\eea
If we give a proper $a=a(t)$ and therefore $H=H(t)$, Eq.~(\ref{YM13}) can be
regarded as a third
order differential equation with respect to $\lambda$. If we find the
solution of $\lambda$
with three constants of the integration, we find the explicit form of
$P(\phi)=P(t)$ by using
(\ref{YM12}) and further obtain $Q(\phi)$ by using (\ref{YM9}). Then we
find the explicit form
of three parameter families of the action (\ref{YM2}).
This tells that almost arbitrary time development of the university could
be realized by the action
(\ref{YM2}) or (\ref{YM1}).

As an example, we may consider the case of the power law expansion:
\be
\label{YM14}
a=\left(\frac{t}{t_1}\right)^{h_1}\quad \left(H=\frac{h_1}{t}\right)\ .
\ee
Here $t_1$ and $h_1$ are constants. By assuming
\be
\label{YM15}
\lambda = \left(h_1 - 1\right) \ln \left(\frac{t}{t_1}\right) +
\lambda_1\ ,
\ee
($\lambda_1$ is a constant), Eq.~(\ref{YM3}) reduces to the algebraic
equation:
\be
\label{YM16}
0=\frac{2h_1}{h_1-1} \bar{X}^2 + \left( -4h_1^2 + 13 h_1 + 2 \right) \bar{X}
+ \left(h_1 -1 \right)^2 \left(h_1 - 2\right)\left(4H_1 -20\right)\ .
\ee
Here $\bar{X}=\bar{\alpha}^2 t_1^2 \e^{2\lambda}$.
If (\ref{YM16}) has a real positive solution
with respect to $\bar{X}$,
we obtain $\lambda_1$ and therefore the exact form of $\lambda$.
Then we can reconstruct a model to give the power expansion (\ref{YM14}).
Similarly, any other universe evolution history maybe reproduced by
specific form of the action under consideration.

\section{Conclusion}

In the present paper, we have considered inflationary cosmology and
the late-time accelerated expansion of the universe in the YM theory, in which
the YM field couples to a function of the scalar curvature, 
in order to investigate the cosmological consequences of the non-minimal 
gravitational coupling of the YM filed. 
As a result,
we have shown that power-law inflation can be realized due to the non-minimal
gravitational coupling of the YM field. 
Moreover, we have demonstrated that both inflation and the late-time
accelerated expansion of the universe can be realized in a modified
YM-$F(R)$ gravity proposed in Ref.~\cite{Nojiri:2007as} 
which is consistent with solar system tests. 
Furthermore, we have shown that this result can be realized also
in a non-minimal vector-$F(R)$ gravity. 
In addition, we have considered the duality of the non-minimal
electromagnetic theory and that of the non-minimal YM theory.
Furthermore, we also discussed the reconstruction of the YM theory from
the known universe history expansion.
As an example, it has been shown that a model to give the power expansion
of the scale factor can be reconstructed.

Finally, we remark the following point.
It is interesting that our models maybe extended by another
gauge-non-invariant non-minimal coupling with the curvature like the ones
done in Ref.~\cite{Dimopoulos:2008rf} 
(for models of vector curvaton, see~\cite{vector curvaton}).
Such non-minimal vector curvaton may give extra contribution to curvature
perturbations if compare with the present models.
Another important point is related with the exit from the inflation.
In the models under consideration it maybe realized via the gravitational
scenario, as the instability of de Sitter universe, due to extra
gravitational terms.
This scenario will be investigated in detail elsewhere. It maybe also
relevant for the study of future universe: if our universe will stay as
$\Lambda$-CDM one forever or it will evolve to other singular/non-singular
state.

\section*{Acknowledgments}
We are grateful to M.~Sasaki for very helpful discussion of
related problems.
The work of K.B. was supported in part by the
open research center project at Kinki University and that by S.D.O. was
supported in part by MEC (Spain) projects FIS2006-02842 and
PIE2007-50/023.
This work by S.N. is supported in part by the Ministry of Education,
Science, Sports and Culture of Japan under grant no.18549001 and 21st
Century COE Program of Nagoya University provided by the Japan Society
for the Promotion of Science (15COEG01).


\end{document}